\documentclass[final,english,twocolumn,amssymb,aps,prb,
longbibliography]{revtex4-2}
\usepackage{graphicx}
\usepackage{natbib}
\usepackage{amsmath}
\usepackage{amssymb}
\usepackage{wasysym}
\usepackage{appendix}
\usepackage{bm}
\usepackage{overpic}
\usepackage[dvipsnames]{xcolor}{\huge }
\usepackage[section]{placeins}
\definecolor{darkblue}{rgb}{0,0,.65}
\definecolor{darkgreen}{rgb}{0.28,0.41,0.19}

\usepackage{multirow}

\usepackage{nicefrac}

\usepackage[%
    pdfauthor={Imre Hagymasi},%
  pdfstartview=FitH,%
  breaklinks=true,%
  bookmarks=true,%
  colorlinks=true,%
  anchorcolor=black,%
  citecolor=blue,
  filecolor=black,%
  menucolor=black,%
  urlcolor=darkblue,%
  linkcolor=blue,%
 ]{hyperref}
\usepackage[all]{hypcap} %

\usepackage[capitalize]{cleveref}

\graphicspath{{images/}}
\begin{document}

\title{A Tale of Two Plateaus: Competing Orders in Spin-1 and Spin-$\tfrac{3}{2}$ Pyrochlore Magnets}

\author{Imre Hagym\'asi}
\affiliation{Institute 
for Solid State
Physics and Optics, Wigner Research Centre for Physics, Budapest H-1525 P.O. 
Box 49, Hungary
}
\date{\today}

\begin{abstract}
We use large-scale density-matrix renormalization group simulations with bond dimensions up to $20\ 000$ to determine the magnetization curves of spin-1 and spin-$\tfrac{3}{2}$ pyrochlore Heisenberg antiferromagnets. Both models exhibit a robust half-magnetization plateau, and we find that the same 16-site state (quadrupled unit cell) is selected in both cases on the largest 64-site cubic cluster we consider for the plateau state. This contrasts sharply with the effective quantum dimer model prediction which favors the ``R'' state, and demonstrates the breakdown of the perturbative mechanism at the Heisenberg point. These results provide a nonperturbative characterization of field-induced phases in pyrochlore magnets and predictive guidance for spin-1 and spin-$\tfrac{3}{2}$ materials.
\end{abstract}

\maketitle

%\section{Introduction}
\par \emph{Introduction.---}
Frustrated quantum magnets provide a fertile setting for unconventional collective phenomena, as lattice geometry can prevent the simultaneous minimization of local exchange energies.
On the three-dimensional pyrochlore lattice of corner-sharing tetrahedra, this frustration gives rise already in the classical limit to an extensively degenerate manifold and emergent Coulomb-phase correlations, as realized for instance in spin-ice materials
\cite{ramirez_zero-point_1999,bramwell_spin_2001,castelnovo_magnetic_2008}.
For classical Heisenberg spins, a similarly degenerate ground-state manifold with dipolar correlations is obtained
\cite{Isakov_dipolar_prl,Henley_dipolar_prb}.
Anisotropies, further interactions are known to lift this degeneracy and induce long-range order
\cite{bramwell_singleionanisotropy_1994,moessner_anisotropy_1998,eljahal_dzyaloshinsky_2005}.
In the quantum regime, even the long-studied spin-$\tfrac{1}{2}$ pyrochlore Heisenberg antiferromagnet, once considered a prime candidate for a three-dimensional quantum spin liquid
\cite{canals_lacroix_prb_2000,CanalsLacroix_prl,iqbal_quantum_2019},
has recently been shown to develop subtle symmetry breaking
\cite{hagymasi_possible_2021,astrakhantsev_broken-symmetry_2021,hering2021dimerization,schaefer2022abundance}.

A magnetic field provides an additional tuning parameter which partially lifts the degeneracy while preserving strong frustration.
In frustrated quantum magnets this often leads to magnetization plateaus and field-induced quantum phases
\cite{honecker_magnetization_2004,richter_quantum_2004,Moessner_kagice,capponi_numerics_kagome_plateaus_2013,Udagawa_kagice}.
In two-dimensional lattices, in particular on the kagome lattice, a wide range of plateau states, magnetization anomalies and field-driven quantum phases have been reported
\cite{honecker_kagome_2005,sakai_kagome_2011,nakano_kagome_2014,nakano_kagome_2018,plat_kagome_2018,chen_kagome_2018,schnack_kagome42_2018,nishimoto_controlling_2013,capponi_numerical_2013,zhitomirsky_prl_2000,coletta_prb_2013}.
In this context, localized-magnon physics provides a quantitative description of magnetization jumps and magnon-crystal phases close to saturation
\cite{zhitomirsky_exact_2004,richter_exact_2004,derzhko_localized_magnons_2007,schmidt_linear_2002,schulenburg_theory_kagome_magnon_groundstate_2002}.

In contrast, the magnetization process of the three-dimensional pyrochlore Heisenberg antiferromagnet remains far less understood \cite{pal_prb_2019,hagymasi_field_2022}, especially for higher spins directly relevant for experiments. The material NaCaNi$_2$F$_7$ realizes an almost ideal spin-1 pyrochlore Heisenberg magnet \cite{plumb_continuum_2019}, while Cr-based spinels such as CdCr$_2$O$_4$ and ZnCr$_2$O$_4$ are good approximations to $S=\tfrac{3}{2}$ Heisenberg pyrochlore magnets. The latter exhibit a pronounced half-magnetization plateau together with several field-induced transitions
\cite{ueda_experiment_2005,kojima_experiment_2010}.
These features are commonly attributed to spin--lattice coupling, which lifts the classical degeneracy and stabilizes collinear states
\cite{penc_prl_2004}.
Importantly, in the \emph{pure} classical Heisenberg model no half-magnetization plateau occurs in the absence of spin--lattice coupling, raising the central question whether quantum fluctuations alone can stabilize such a plateau.

\begin{figure}[t]
    \vspace{1cm}
    \centering
    \begin{tabular}{cc}
        \shortstack{``$S$'' state\\[2mm]
            \includegraphics[width=0.49\columnwidth]{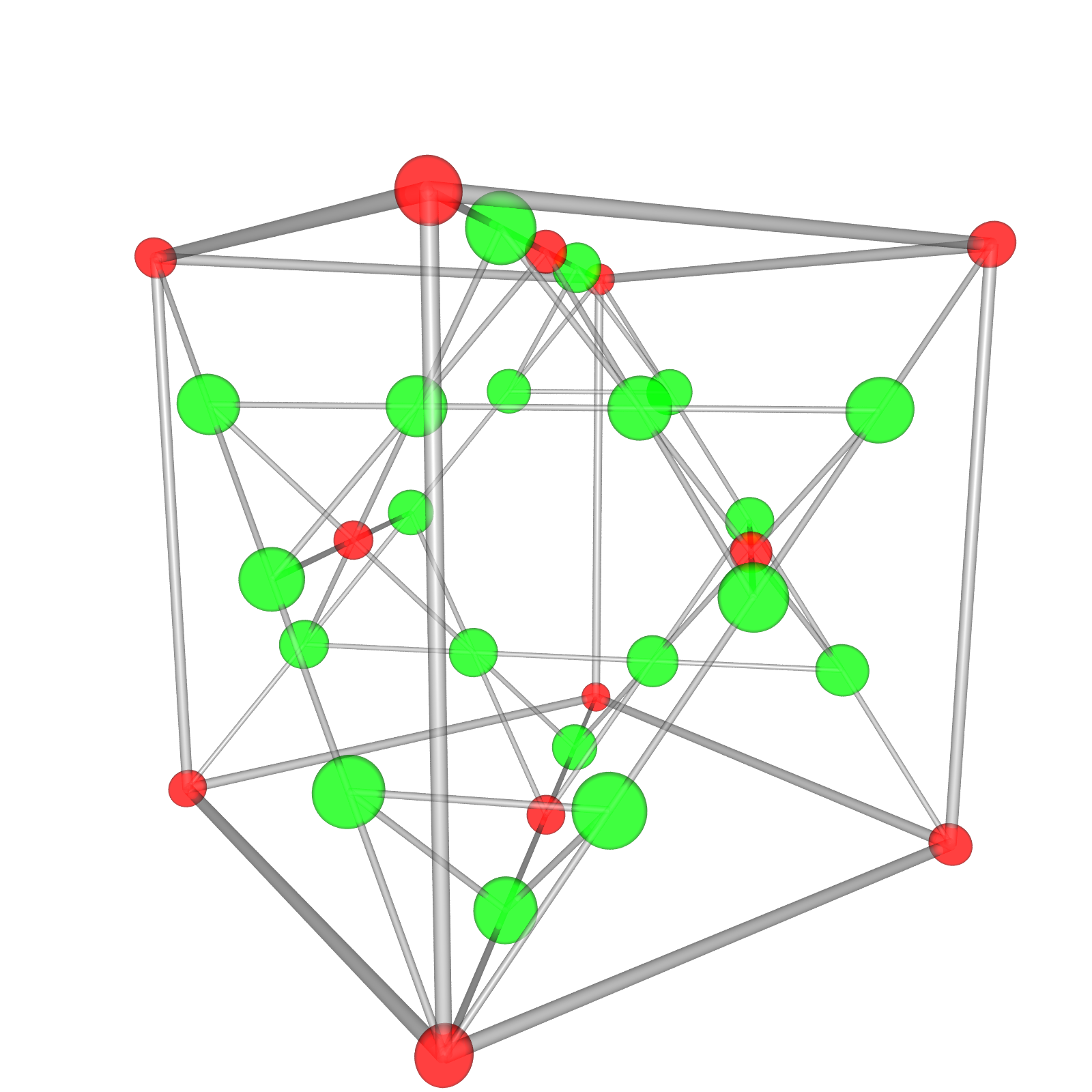}}
        &
        \shortstack{``$R$'' state\\[2mm]
            \includegraphics[width=0.49\columnwidth]{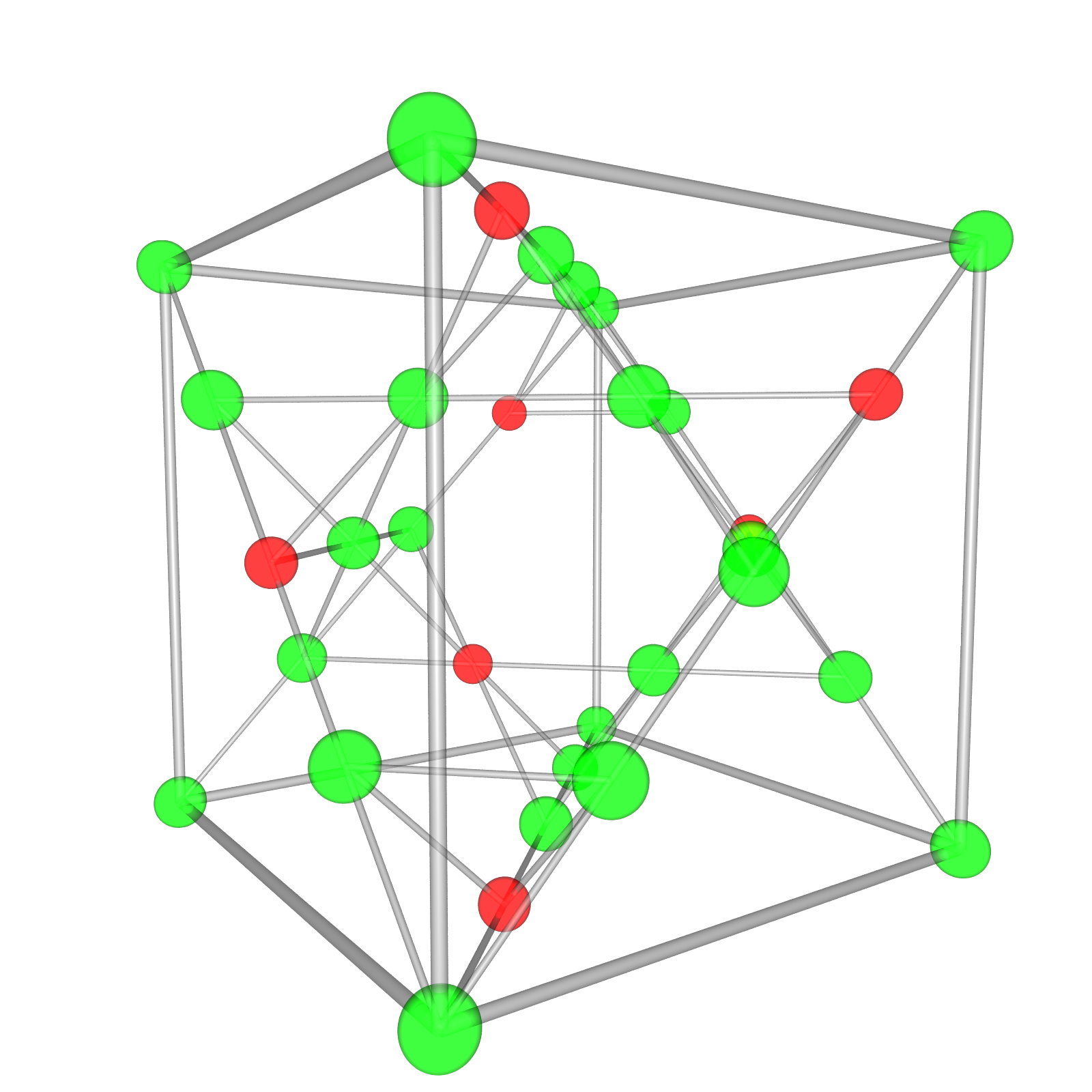}}
    \end{tabular}
    \caption{On-site magnetization patterns on the $m/m_{\rm sat}=\tfrac{1}{2}$ plateau for the cubic $N=64$ cluster. Only one cubic unit cell is shown. Red spheres mark minority-spin sites; sphere radii are proportional to $|\langle S_i^z\rangle|$. For the ``S'' state we find $\langle S_i^z\rangle \sim 0.88$ (majority) and $\sim -0.64$ (minority) for $S=1$, and $\langle S_i^z\rangle \sim 1.36$ (majority) and $\sim -1.09$ (minority) for $S=\tfrac{3}{2}$ at bond dimension $\chi=12\,000$. The corresponding values for the ``R'' state are essentially the same.}
    \label{fig:magnetization}
\end{figure}

From a theoretical perspective, a prominent description of the half-magnetization plateau is provided by the perturbative analysis of Bergman \textit{et al.} \cite{bergman_prl_2006}, who derived an effective quantum dimer model acting within the three-up--one-down ($uuud$) manifold of the spin-$\tfrac{3}{2}$ model in a strong magnetic field and identified a particular ordered plateau state, the so-called ``R'' state (see Fig.~\ref{fig:magnetization}), selected by an order-by-disorder mechanism
\cite{bergman_prl_2006}.
Subsequent work demonstrated that several symmetry-distinct plateau states are in principle allowed and the perturbative hierarchy of effective couplings collapses upon approaching the isotropic Heisenberg point
\cite{bergman_prb_2006,bergman_prb_2007}.
As a result, it remains unresolved whether the ``R'' state or another candidate ($U(1)$ spin liquid or a resonating plaquette state) are realized in the Heisenberg limit for $S=1$ and $S=\tfrac{3}{2}$.

Here, we perform large-scale density-matrix renormalization group (DMRG) simulations of the spin-1 and spin-$\tfrac{3}{2}$ pyrochlore Heisenberg antiferromagnets in a magnetic field.
We show that both models exhibit a robust half-magnetization plateau in the Heisenberg limit.
While the perturbative ``R'' state is stabilized for $S=\tfrac{3}{2}$ on the smallest cubic cluster, we find that on the largest cubic clusters we considered, both spin-1 and spin-$\tfrac{3}{2}$ systems favor a distinct ordered plateau state with a 16-site magnetic unit cell, referred to as the ``S'' state.
Our results demonstrate the failure of the perturbative plateau scenario close to the Heisenberg point and establish the nonperturbative selection of the half-plateau state for experimentally relevant spin lengths.

\par \emph{Model and method.---}
We study the spin-$S$ Heisenberg antiferromagnet on the pyrochlore lattice in an external magnetic field, $h$:
\begin{equation}
H = J \sum_{\langle i,j\rangle} \bm{S}_i\!\cdot\!\bm{S}_j - h \sum_i S_i^z ,
\end{equation}
where $\bm{S}_i= (S_i^x,\,S_i^y,\,S_i^z)^T$ are spin operators acting on site $i$ and the summation extends over the nearest-neighbor bonds of the pyrochlore lattice.
The ground states are obtained using large-scale DMRG calculations with explicit SU(2) and U(1) spin symmetries
\cite{hubig:_syten_toolk,hubig17:_symmet_protec_tensor_networ,hubig_2015,McCulloch_2007}.
Although DMRG is intrinsically a one-dimensional method \cite{white_1992,white_1993,noack2005,schollwock_review_2011,hallberg_review}, it has been successfully applied to two- and three-dimensional frustrated magnets by mapping the lattice onto a one-dimensional snake path
\cite{white_2d_dmrg,ummethum_numerics_2013,hagymasi_prl_2021}.

We consider periodic clusters of up to $N=128$ sites and keep bond dimensions of up to $20\,000$ SU(2) or U(1) states.
To reliably compare competing low-energy states, we extrapolate the variational energies to the error-free limit using the two-site variance as a convergence measure \cite{hubig_prb_2018}.
For each candidate ordering pattern we perform independent simulations starting from several symmetry-related initial states and retain the lowest-energy extrapolation.
Additional details of the simulation and the periodic clusters are provided in the End Matter.

\par
\begin{figure}[!ht]
%\centering
\includegraphics[width=\columnwidth]{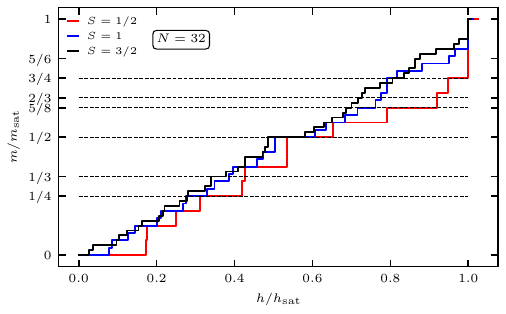}
\includegraphics[width=\columnwidth]{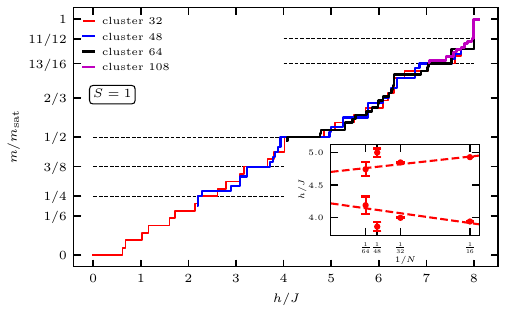}
\caption{ Normalized magnetization curves of pyrochlore clusters in a magnetic field. 
Upper panel: $m/m_{\rm sat}$ versus $h/h_{\rm sat}$ for the cubic $N=32$ cluster and spin lengths $S=\tfrac{1}{2},1,\tfrac{3}{2}$, where $m=\langle\psi_0|S^z_{\rm tot}|\psi_0\rangle$, $m_{\rm sat}=NS$, and $h_{\rm sat}=8JS$. 
Lower panel: magnetization for the $S=1$ model on clusters up to $N=108$; for $N=48$ and $N=64$ only the upper half of the curve is reliably accessible, while for $N=108$ only the high-field regime is shown. 
Inset: finite-size extrapolation of the lower and upper critical fields of the $m/m_{\rm sat}=\tfrac{1}{2}$ plateau, yielding $h_- = 4.22(1)J$ and $h_+ = 4.70(1)J$ in the thermodynamic limit; the fit is restricted to cubic clusters and includes the primitive 16-site cubic unit cell.
}
\label{fig:magnetization_curve_00}
\end{figure}

 \emph{Magnetization curve.---}
We determine the field dependence of the ground-state magnetization following the same protocol as in Ref.~\cite{hagymasi_field_2022}.
For the cubic 32-site cluster we obtain the full magnetization curves for $S=1$ and $S=\tfrac{3}{2}$, shown in the upper panel of Fig.~\ref{fig:magnetization_curve_00} together with the corresponding $S=\tfrac{1}{2}$ data from Ref.~\cite{hagymasi_field_2022}.
The lower panel of Fig.~\ref{fig:magnetization_curve_00} displays the magnetization curves for the spin-$1$ model on larger clusters.

Two prominent features emerge already on the 32-site cluster.
First, a magnetization jump appears close to saturation, whose magnitude decreases with increasing spin length.
This jump is attributed to the condensation of localized magnons \cite{zhitomirsky_exact_2004,richter_exact_2004,hagymasi_field_2022,derzhko_localized_magnons_2007}.
In the thermodynamic limit, each localized magnon occupies a 12-site motif \cite{hagymasi_field_2022}, implying magnon-crystal plateaus at $m/m_{\rm sat}=11/12$ for $S=1$ and $m/m_{\rm sat}=17/18$ for $S=\tfrac{3}{2}$.
Consistent with this picture, the localized-magnon plateau is commensurate with the 108-site cluster for $S=1$, but remains extremely narrow and is unlikely to retain a finite width in the thermodynamic limit.

Second, a robust half-magnetization plateau is observed for all three spin lengths on the 32-site cluster.
For $S=1$, this plateau persists on larger clusters and allows for a finite-size extrapolation of its critical fields using only cubic clusters, which confirms a finite plateau width in the thermodynamic limit (see inset of Fig.~\ref{fig:magnetization_curve_00}).
In the following, we focus on the internal ordering of the half-magnetization plateaus for $S=1$ and $\tfrac{3}{2}$.

Finally, we briefly comment on the zero-field properties.
Although the DMRG findings for the ground states of the spin-$\tfrac{1}{2}$ and spin-$1$ models have been discussed previously \cite{hagymasi2022prb,hagymasi_prb_2024}, we note that for $S=\tfrac{3}{2}$ on the 32-site cluster we obtain a ground-state energy of $E/N=-3.06(5)J$ and a triplet gap of $0.31(5)J$.
The corresponding ground state exhibits the same dimensional reduction of magnetic correlations as observed in the spin-$1$ case \cite{hagymasi2022prb,hagymasi_prb_2024}.

\par
\begin{figure}[!t]
%\centering
\includegraphics[width=\columnwidth]{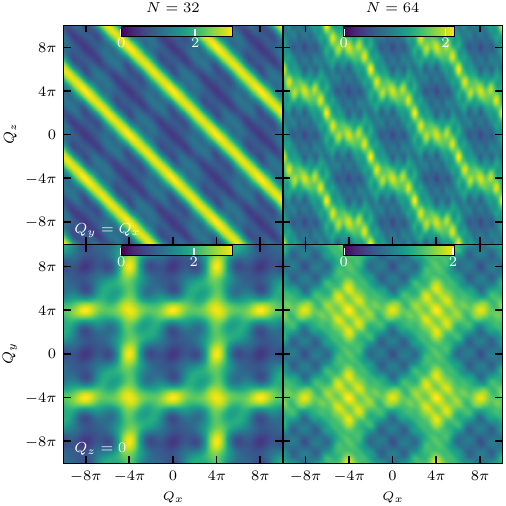}
\caption{Spin structure factor of the spin-1 model on different clusters at the half-magnetization plateau, computed at bond dimension $\chi=10\,000$. The upper (lower) panels show cuts in the $[hhl]$ ($[hl0]$) plane. The appearance of line-like features signals dimensional reduction and the breaking of threefold rotational symmetry, whose form depends on the cluster geometry.
}
\label{fig:structure_factor_spin1}
\end{figure}

 \emph{Ordering on the $m/m_{\rm sat}=\tfrac{1}{2}$ plateau.---}
We first discuss the structure of the half-magnetization plateau in the spin-1 model.
On the cubic 32-site cluster, we find a state with strongly polarized triangular planes and weakly polarized kagome planes, resulting in $\langle\sum_{i\in{\rm tet}}S_i^z\rangle\simeq2$ per tetrahedron.
This state breaks the $C_3$ rotational symmetry of the pyrochlore lattice, as is also visible in the corresponding structure factor (shown in Fig.~\ref{fig:structure_factor_spin1}) defined as $  S(\vec{Q})= \tfrac{1}{N} \sum_{i j} \langle \vec{S}_i\cdot 
\vec{S}_j\rangle_c \cos\left[\vec{Q}\cdot 
\left(\vec{R}_i-\vec{R}_j\right)\right]$, where $\vec{R}_i$ are the real-space coordinates of sites and the index $c$
denotes the connected part of the correlation matrix.
On the 48-site clusters, a related form of dimensional reduction is observed, where the plateau state is characterized by non-intersecting strongly and weakly polarized chains.

We next turn to the cubic 64-site cluster, which allows us to compare several symmetry-distinct ordering patterns.
Starting from random initial states, we find that all low-energy plateau states obey the $uuud$ constraint but differ in the spatial arrangement of the minority spins, indicating a quantum order-by-disorder selection mechanism.
Guided by the ordering tendencies found on smaller clusters, and by perturbative expectations, we focus on three candidate states.
The first is the simplest $\boldsymbol{q}=0$ plateau state with the primitive four-site unit cell, in which the same spin is flipped on every tetrahedron, leading to a dimensional reduction into alternating ferromagnetic and antiferromagnetic chains.
The second candidate is the previously proposed ``R'' state.
The third candidate is a distinct state with an enlarged 16-site magnetic unit cell, referred to as the ``S'' state (Fig.~\ref{fig:magnetization}), in which the minority spins are arranged in a staggered pattern within the unit cell.
This shifted arrangement produces non-intersecting antiferromagnetic lines confined to kagome planes and represents a qualitatively different realization of dimensional reduction.

\begin{figure}[!ht]
%\centering
\includegraphics[width=\columnwidth]{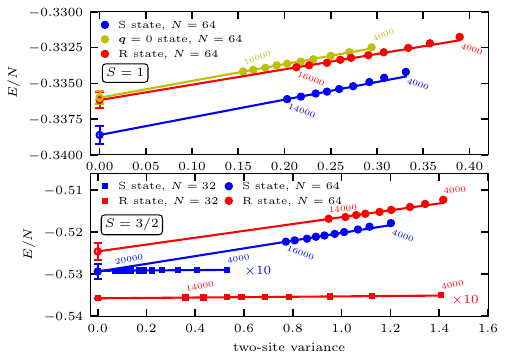}
\caption{Variance-based extrapolation of the energy per site for different plateau candidates. The upper panel shows results for the spin-1 model on the cubic 64-site cluster, while the lower panel presents data for the spin-$\tfrac{3}{2}$ model on the 32- and 64-site clusters. For clarity, variance values for the 32-site cluster are rescaled by a factor of 10. Error bars are defined as one quarter of the difference between the lowest-variance data point and the extrapolated value. Numbers next to the symbols indicate the bond dimension $\chi$. For spin-$\tfrac{3}{2}$, data for the $\boldsymbol{q}=0$ state are omitted for clarity, as its energies lie above those of the other candidates.
 }
\label{fig:extrapolation}
\end{figure}

To determine the energetically preferred ordering, we extrapolate the variational energies of these candidate states to the error-free limit using the two-site variance.
The results are summarized in Fig.~\ref{fig:extrapolation} and Table~\ref{tab:energies}.
\begin{table}[!t]
	\centering
	\begin{tabular}{l@{\hspace{5mm}}c@{\hspace{5mm}}c@{\hspace{5mm}}c}
		\hline\hline
		  & ``S'' state & ``R'' state & $\boldsymbol{q}=0$ state   \\
		\hline
		$S=1$   &  $-0.3386(6)$ & $-0.3361(6)$ & $-0.3360(5)$   \\
		$S=\tfrac{3}{2}$   &  $-0.529(2)$ & $-0.524(2)$ & $-0.523(2)$ \\[0.05cm]
		\hline
	\end{tabular}
	\caption{Extrapolated ground-state energies per site (in units of $J$) for different half-magnetization plateau candidates on the cubic 64-site cluster. Energies are obtained from variance-based extrapolations, with uncertainties reflecting the extrapolation error. For both spin lengths, the ``S'' state yields the lowest energy among the considered candidates.}
	\label{tab:energies}
\end{table}
For the spin-1 model on the cubic 64-site cluster, the ``S'' state consistently yields the lowest extrapolated energy, while both the ``R'' and the $\boldsymbol{q}=0$ states remain higher in energy.
The structure factor of the ``S'' state is shown in Fig.~\ref{fig:structure_factor_spin1}.

Next,  the spin-$\tfrac{3}{2}$ case is considered.
On the cubic 32-site cluster, the ``R'' state yields the lowest extrapolated energy,
$E_{\rm ext}^R(N=32)/N=-0.5357J$, compared to
$E_{\rm ext}^S(N=32)/N=-0.5292J$,
consistent with perturbative expectations.
However, on the cubic 64-site cluster the situation reverses.
While the extrapolated energy of the ``S'' state remains essentially unchanged, the energy of the ``R'' state increases, and the ``S'' state becomes energetically favored (Table~\ref{tab:energies}).
The $\boldsymbol{q}=0$ state also extrapolates to a higher energy.
Since the cubic 64-site cluster preserves the full lattice symmetry and allows for the most accurate energy extrapolations accessible in our simulations, we conclude that the ``S'' state is favored over the ``R'' state for spin-$\tfrac{3}{2}$ as well.
The corresponding structure factors are shown in Fig.~\ref{fig:structure_factor_spin1.5} of the End Matter.
Although the ``R'' state breaks $C_3$ rotational symmetry, its structure factor on the 32-site cluster exhibits an almost symmetric signal in the $[hhl]$ plane, in contrast to the pronounced line features observed for the ``S'' state, which closely resemble those found in the spin-1 model.
\par As a closing remark we mention that exploratory simulations on a larger cubic, $N=128$, cluster up to a bond dimension of $8000$, although substantially less converged, consistently yield a lower variational energy for the ``S'' state than for the ``R'' state at comparable variance for both spin lengths (see End Matter for further details), supporting the stability of the ``S'' state on larger length scales.
\par \emph{Discussion.---}
It is worth putting our results into context in light of the perturbative descriptions of the half-magnetization plateau based on an effective quantum dimer model (QDM) acting within the $uuud$ manifold
\cite{bergman_prl_2006,bergman_prb_2006,bergman_prb_2007}.
In the easy-axis and strong-field limit the leading processes are generated on hexagonal loops of the pyrochlore lattice,
\begin{align}
H_{\rm QDM}
=
- K \sum_{\hexagon}
\bigl(
|\hexagon_A\rangle\langle\hexagon_B|+\text{h.c.}
\bigr)
+\\
+V \sum_{\hexagon}
\bigl(
|\hexagon_A\rangle\langle\hexagon_A|
+
|\hexagon_B\rangle\langle\hexagon_B|
\bigr),
\end{align}
where $\hexagon_{A/B}$ denote the two flippable hexagon configurations, $V\sim\Delta^{6}$ and $K\sim\Delta^{6S}$, where $\Delta$ denotes the ratio of transverse and Ising couplings.
For $S=\tfrac{3}{2}$, the leading-order analysis yields $|V|\gg|K|$ and $V<0$, favoring the maximally flippable ``R'' state in the simplified QDM \cite{bergman_prl_2006,sikora_prb_2011}.

However, the extended perturbative treatments show that additional diagonal contributions arising from \emph{non-flippable} hexagons and longer virtual processes substantially modify the effective Hamiltonian
\cite{bergman_prb_2006,bergman_prb_2007}.
As a result, the ``R'' state is no longer generically selected uniquely by the diagonal part alone, and several competing phases become allowed, including resonating plaquette states and a $U(1)$ spin-liquid scenario. Moreover, upon extrapolation toward the isotropic Heisenberg limit, the hierarchy of effective couplings collapses and kinetic and potential terms become comparable, rendering the perturbative expansion quantitatively uncontrolled.

Our nonperturbative DMRG results are consistent with this picture.
While the $uuud$ constraint is robustly satisfied and thus the mapping to a dimer description remains qualitatively meaningful, the minimal hexagon-only QDM is insufficient to capture the energetic competition at the Heisenberg point.
In particular, the stabilization of the ``S'' state for both $S=1$ and $S=\tfrac{3}{2}$ on the cubic 64-site cluster indicates that higher-order loop processes and diagonal contributions beyond the flippable-hexagon sector play an essential role in determining the plateau order.
This provides direct evidence that the perturbative scenarios break down in the Heisenberg limit and that the selection of the half-plateau state is governed by genuinely nonperturbative quantum fluctuations.
\par At last, we briefly mention that the half-plateau terminates asymmetrically for the clusters we studied, with a discontinuity on the low-field side, reminiscent of the asymmetric plateau termination demonstrated very recently for quantum kagome ice \cite{udagawa2025spinonbandflatteningemergent}, where different quasiparticles control the approach from below and above.
\par \emph{Conclusions.---}
Using large-scale DMRG, we have investigated the magnetization process of the spin-$1$ and spin-$\tfrac{3}{2}$ pyrochlore Heisenberg antiferromagnets in a magnetic field.
Both models exhibit a pronounced half-magnetization plateau.
For the spin-$1$ case, simulations on several cluster sizes allow us to establish a finite plateau width in the thermodynamic limit, despite the absence of such a plateau in the classical Heisenberg model without spin--lattice coupling.

A central result of this work is the identification of the ordering pattern on the half plateau.
On the largest cubic cluster, for which the extrapolation can be performed accurately, both spin-$1$ and spin-$\tfrac{3}{2}$ systems stabilize the same plateau state with a 16-site magnetic unit cell (the ``S'' state), characterized by a staggered arrangement of minority spins and a dimensional reduction of magnetic correlations.
For spin-$\tfrac{3}{2}$, we find that the previously proposed ``R'' state is favored only on the smaller 32-site cluster, while it is overtaken by the ``S'' state on the cubic 64-site cluster.

These findings demonstrate that the perturbative scenarios are no longer valid at the isotropic Heisenberg point.
Although the local $uuud$ constraint remains robust, the selection of the plateau order is controlled by nonperturbative quantum fluctuations and requires effective processes beyond the simplest hexagon-flip description.
Our results thus provide direct numerical evidence that the half-magnetization plateau of the pyrochlore Heisenberg antiferromagnet is governed by a qualitatively different mechanism than suggested by leading-order perturbation theory.

\par \emph{Acknowledgements.---} The author thanks Yasir Iqbal, Roderich Moessner and Johannes Reuther for useful comments and for careful reading of the manuscript. The author also acknowledges fruitful discussions with Masayuki Hagiwara, Karlo Penc, Kemp Plumb and Robin Sch\"afer. This project was supported by the Hungarian National Research, Development   and   Innovation Office (NKFIH) through Grant No.~FK142985 as well as by the J\'anos Bolyai Research Scholarship of the Hungarian Academy of Sciences.
\bibliography{pyrochlore_field,pyrochlore,pyrochlore2,pyrochlore2trim}
\section*{End Matter: Numerical details and additional analyses}

\subsection*{Clusters and boundary conditions}
The pyrochlore lattice may be viewed as an fcc Bravais lattice decorated with a four-site tetrahedral basis. We adopt the primitive fcc vectors $\bm{a}_1=\tfrac{1}{2}(1,1,0)^T$, $\bm{a}_2=\tfrac{1}{2}(1,0,1)^T$, and $\bm{a}_3=\tfrac{1}{2}(0,1,1)^T$, and assign to each Bravais point the basis positions $\bm{b}_0=\bm{0}$ and $\bm{b}_i=\tfrac{1}{2}\bm{a}_i$ for $i=1,2,3$. Every lattice site can then be written as $\bm{R}_{\alpha,n_1,n_2,n_3}=n_1\bm{a}_1+n_2\bm{a}_2+n_3\bm{a}_3+\bm{b}_\alpha$, where $n_1,n_2,n_3\in\mathbb{Z}$ and $\alpha\in\{0,1,2,3\}$ labels the basis.
\par We consider periodic clusters compatible with the pyrochlore lattice geometry.
In the main text, we focus on cubic clusters whenever possible, since they preserve the full point-group symmetry of the underlying fcc lattice and admit the longest noncontractible winding loops accessible in our simulations. In particular, the cubic 64-site cluster is used as the primary reference geometry for the identification of the plateau ordering patterns.
Additional clusters with 16, 32, 48, 108 and 128 sites are used to assess finite-size trends of the magnetization curve or to test the stability of the observed plateau states.
For the 48-site clusters, to test the robustness of dimensional reduction, several inequivalent shapes (\emph{cf.} the 48-site clusters in \cite{hagymasi_possible_2021}) were considered. The frame vectors of the clusters that are used in the paper are listed in Table \ref{tab:clusters}.
\begin{table}[!h]
	\centering
	\begin{tabular}{l@{\hspace{5mm}}c@{\hspace{5mm}}c@{\hspace{5mm}}c}
		\hline\hline
		cluster & $\bm{c}_1$ &  $\bm{c}_2$ & $\bm{c}_3$   \\
		\hline
        16  &  $(1,0,0)^T$ & $(0,1,0)^T$ & $(0,0,1)^T$ \\
		32   &  $2 \bm{a}_1$ & $2 \bm{a}_2$ & $2\bm{a}_3$   \\
        48   &  $2 \bm{a}_1$ & $2 \bm{a}_2$ & $3\bm{a}_3$   \\
		64   &  $(1,1,1)^T$ & $(1 ,1,-1)^T$ & $(-1,1,-1)^T$ \\
        108   & $3 \bm{a}_1$ & $3 \bm{a}_2$ & $3\bm{a}_3$   \\
        128  &  $(2,0,0)^T$ & $(0,2,0)^T$ & $(0,0,2)^T$ \\
		\hline
	\end{tabular}
	\caption{Frame vectors $\bm{c}_1$, $\bm{c}_2$, $\bm{c}_3$ of the clusters used in this work.  The clusters (except for the 48-site one) respect all point symmetries of the fcc lattice.}
	\label{tab:clusters}
\end{table}

\subsection*{Details of the DMRG simulation for the half-plateau states}

For clusters beyond 48 sites, straightforward DMRG optimizations starting from random initial states frequently converge to metastable low-energy configurations.
To overcome this problem, we employ weak pinning fields during the initial sweeps in order to stabilize targeted ordering patterns.
After convergence, the pinning fields are removed and the states are further optimized without bias.
Importantly, we find that the pinning fields can already be switched off at very small bond dimensions ($\chi \sim 30$), after which the states remain stable and converge smoothly to the same ordered patterns.
This demonstrates that the resulting states are intrinsic variational solutions of the Hamiltonian rather than artifacts of the pinning fields.
Overall, this procedure results in systematically lower variational energies and significantly improved convergence compared to simulations initialized from random states.
\par For each candidate ordering pattern on the half-magnetization plateau (the ``S'' state, the ``R'' state, and the $\boldsymbol{q}=0$ state), several symmetry-related realizations exist, corresponding to different choices of the minority spin within a tetrahedron. Due to the one-dimensional mapping imposed by the DMRG snake path, these symmetry-related partner states are not represented with identical numerical accuracy at a fixed bond dimension. In order to avoid a systematic bias, we therefore perform independent simulations for all symmetry-related realizations of each candidate state and retain, for the final extrapolation, those realizations that yield the lowest variational energy and smallest variance (even though the partner states converge to the same energy).
\par To extrapolate the variational energies to the error-free limit, we employ the two-site variance \cite{hubig_prb_2018}.
\begin{figure}[!h]
%\centering
\includegraphics[width=\columnwidth]{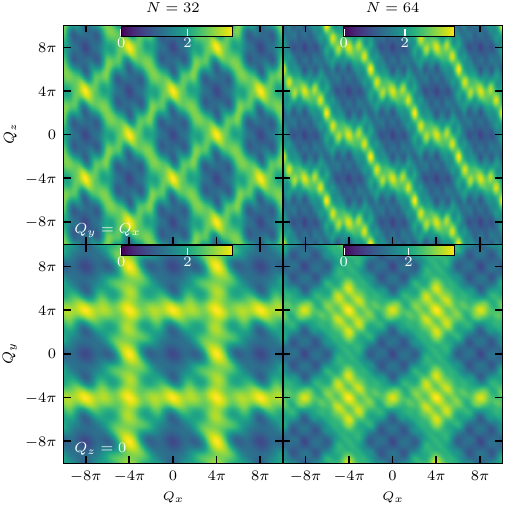}
\caption{Spin structure factor of the spin-$\tfrac{3}{2}$ model at the half-magnetization plateau for different cluster sizes, computed at bond dimension $\chi=10\,000$. The upper (lower) panels show cuts in the $[hhl]$ ($[hl0]$) plane. While the ``R'' state breaks threefold rotational symmetry, its structure-factor signature on the 32-site cluster appears nearly symmetric in the $[hhl]$ plane, in contrast to the pronounced line features observed for the ``S'' state.
}
\label{fig:structure_factor_spin1.5}
\end{figure}
This quantity measures the local deviation from the Schrödinger equation on neighboring site pairs along the DMRG snake path and thus provides a state-local indicator of convergence. For all candidate plateau states shown in the main text, we observe a clear linear regime of the energy as a function of the two-site variance at sufficiently large bond dimensions allowing reliable extrapolation to the error-free limit.
The extrapolation error is estimated as one quarter of the difference between the lowest-variance data point and the extrapolated value.
Because the representation accuracy depends on the snake path and on the detailed structure of the state, comparisons between different plateau candidates are carried out only after extrapolation and not at fixed bond dimension.
\par For the half-magnetization plateau of the 64-site cluster, we find that all low-energy states obtained in our simulations satisfy the local $uuud$ constraint.
This indicates that the restriction to the corresponding constrained manifold remains appropriate even in the isotropic Heisenberg limit. (We not at this point that even for the ``R'' state, evaluating the variational energy of the resulting quantum dimer model is highly nontrivial due to the strongly correlated nature of the resonating hexagons, limiting the quantitative predictive power of the perturbative approach.)
However, the energetic ordering among different plateau states is strongly affected by finite-size geometry.
In particular, the relative stability of the ``R'' and ``S'' states depends sensitively on the availability of long winding loops and on the preservation of lattice symmetries, which motivates our emphasis on the cubic 64-site cluster in the main text.
To further test the scenario presented there, we performed exploratory simulations on a larger cubic 128-site cluster.
Although these calculations are substantially less converged than for $N=64$, they consistently favor the ``S'' state over the ``R'' state at comparable two-site variance (see Table \ref{tab:energies_128}),
\begin{table}[!t]
	\centering
	\begin{tabular}{l@{\hspace{5mm}}c@{\hspace{5mm}}c}
		\hline\hline
		  & ``S'' state & ``R'' state    \\
		\hline
		$S=1$   &  $-0.319$ & $-0.317$    \\
		$S=\tfrac{3}{2}$   &  $-0.487$ & $-0.484$\\[0.05cm]
		\hline
	\end{tabular}
	\caption{Variational ground-state energies per site (in units of $J$) for different half-magnetization plateau candidates on the cubic 128-site cluster obtained at the bond dimension $\chi=8000$.}
	\label{tab:energies_128}
\end{table}
in agreement with the energetics obtained for the 64-site cluster.
Since the 128-site cluster supports longer winding loops (length 8) than the 64-site cluster (length 6), the persistence of the energetic preference for the ``S'' state demonstrates that it is robust against finite-size effects and not tied to a particular cluster geometry.
\par Finally, we mention that a symmetry analysis of the ``S'' state pattern yields the orthorhombic space group $Pmna$, implying cubic-to-orthorhombic symmetry lowering with preserved inversion and broken $C_3$ rotations, while the ``R'' state transforms according to the chiral cubic space group $P4_332$ \cite{bergman_prl_2006}, which lacks inversion symmetry but preserves the full cubic rotational symmetry. The structure factors for the half plateau of the $S=\tfrac{3}{2}$ model are shown in Fig.~\ref{fig:structure_factor_spin1.5} for different system sizes.
\end{document}